# MODEL OF OPINION SPREADING IN SOCIAL NETWORKS

**Igor Kanovsky**
Emek Yezreel College
igork@yvc.ac.il

**Omer Yaari**
Emek Yezreel College
omeryaari@gmail.com

**Keywords:** opinion spreading, social networks, data mining, influencers.

## INTRODUCTION

A computer mediated social network is a modern information system where data is generated, distributed, evaluated and accumulated. It is important to understand how the topology of the network affects the information cycle and to develop network algorithms accordingly. One of the intriguing processes in social networks is information and opinion spreading between actors. This paper attempts to define a proper model of opinion spreading between actors in social networks.

The main approaches for modeling information and opinion spreading is the contagion approach (Kitsak at al. 2010), which is based on the spreading of disease. It suggests that if a healthy person will encounter a sick person, there is a specific probability that the healthy person will get infected. The opinion spreading is just like the disease: if a person without a specific opinion about a topic (not opinioned person) will encounter an opinioned person, the first person will, with some probability, be opinioned.

Recently was proposed another approach: the threshold model or complex contagions (Centola & Macy 2007). The model assumes that a probability of an actor to get opinioned is a sigmoid function of proportion of the actor's opinioned neighbors to total number of neighbors.

Both models contradict the opinion spreading mechanism as it is viewed in sociology (Kleinberg 2008; Centola & Macy 2007).

In our study, we distinguish between two types of information entities. One entity is un-debatable information. For example, facts, information or disease. When one is exposed to activated person, one may be infected, but the next exposure has approximately the same probability for infection. The second type is debatable information: when someone is exposed he or she can accept or choose not to accept the opinion. Examples include consumer tastes, ideas, decisions and so on. The contagion approach is relevant only to the un-debatable information, since in the case of debatable information, conformity is crucial.

In addition, in both approaches mentioned above, there isn't a significant difference in the opinion spreading time depending on which actors act as starting points (Watts & Dodds 2007). This is in contrast with sociology theories according to which there are key actors for opinion spreading in social environments (Katz & Lazarsfeld 1955).

For these reasons we proposed a new model, which captures the main difference between information and opinion spreading.

## OPINION SPREADING

### Model

The model is based on the assumption that the probability of a person to obtain an opinion is a function of the numbers of influencers that the person is encountered with. Jon Kleinberg (2008) called this the "0-1-2" effect, "in which the probability of joining an activity when two friends has done so is significantly more than the twice of the probability of joining when only one has done so".

According to this we need to introduce two different probabilities, one if the person is encountered with one infector, and another probability when encountered with two infectors. For simplifying the situation, we assumed that the opinion may be expressed by Boolean value and there is a time interval on which each person can be exposed to an opinion form two of his friends. We suggested the following model of opinion spreading:

In each time interval for each not opinioned actor in the network we randomly select two of his friends:

    If the two selected friends are not opinioned – the actor stay not opinioned.
    If one is opinioned- the actor gets the opinion in $p_1$ probability.
    If two are opinioned- the actor gets the opinion in $p_2$ probability.

In the case of opinion spreading with 0-1-2 effect $p_2 \gg p_1$, without 0-1-2 effect $p_2 \sim 2 \cdot p_1$.

**Simulations**

Opinion spreading was simulated on different real world social networks (network of e-mail contacts (Leskovec, Kleinberg & Faloutsos 2005) and network of scientific citation (Newman 2001) and social network of user community of tech news site (Leskovec at al. 2008)). Social networks, including the above datasets, obey the "Small World" properties and have power law distribution of actor degree. To distinguish the role of power law degree distribution from the role of Small World, we consider additional free scale networks obtained from the source dataset by randomization the links with preserving nodes degrees.

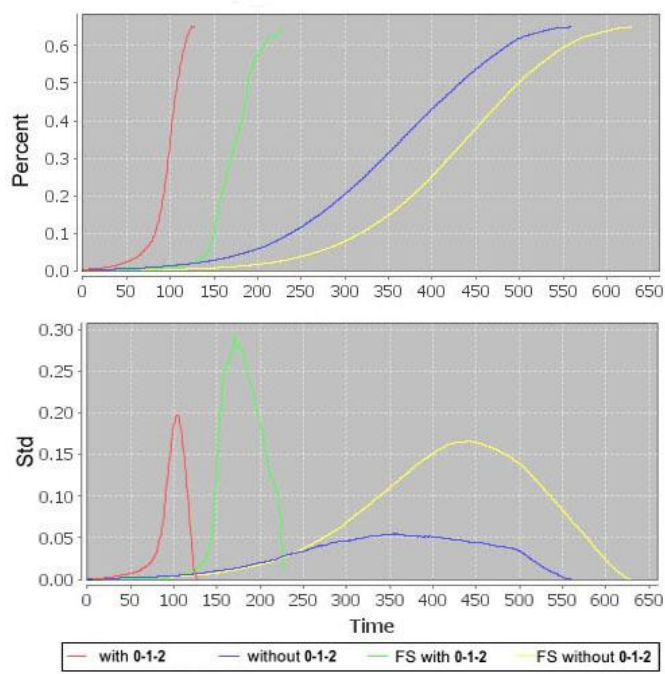

Fig.1. Time dependence of average opinioned actors number and its standard deviation. Enron e-mail contacts network (Leskovec, Kleinberg & Faloutsos 2005). Actors: 36692, edges: 367662, original clustering coefficient: 0.4970. FS – free scale network obtained by original network randomization. $p_1$=0.05, $p_2$=0.9. For 5 simulations starting with the same actor.

For simulation we defined a starting actor whom which will influence the network. In each time iteration according to the model probability outcome some not opinioned actor in the graph became opinioned. The spreading iterates until it reach target fraction of opinioned actor in the graph. Each simulation process was repeated until the standard deviation was low.

The average number of actors with opinion by time line was measured. Typical time dependence is presented on Fig.1. The behavior of the spreading is characterized by a slow incline, until reaching a critical point or tipping point $t_p$ on time line. For accurate definition, we assumed that $t_p$ is reached when 10% of the actors are infected.

Simulation shows that after reaching $t_p$ the spreading speed is dramatically increases. Without the 0-1-2 effect the form of time dependence curve is similar to the one with 0-1-2 effect, however $t_p$ value is significantly larger.

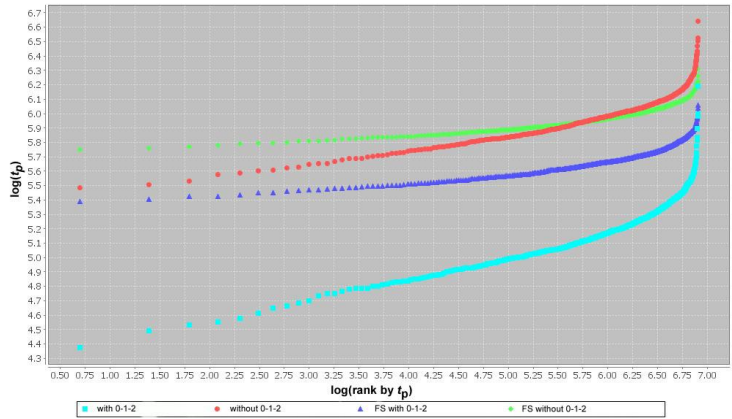

Fig.2. $t_p$ for the actors with high in degree. Enron e-mail contacts network.

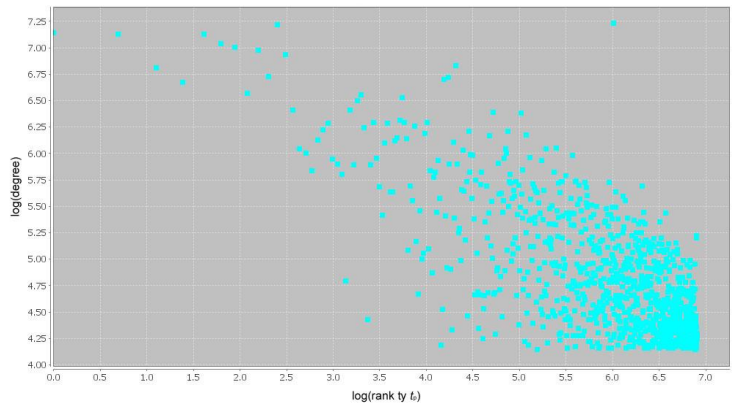

Fig. 3. Actors' degree vs. rank by $t_p$. Enron e-mail contacts network.

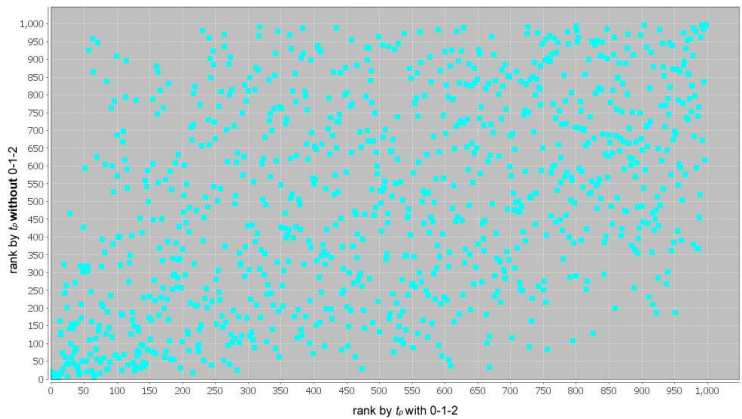

Fig. 4. Actors rank by $t_p$ in the case of 0-1-2 effect vs. rank without 0-1-2 effect.

**Results**

Main interest in the simulations is to understand if there's a difference between $t_p$ for different starting actors of the network. On Fig.2 depicted $log(t_p)$ ordered by increasing value for actors with high degree. Simulations show (see Fig.2) that for 0-1-2 effect the difference is significant, while without 0-1-2 the difference is small. In addition, in randomized scale free network with 0-1-2 effect the difference is even smaller.

The actors with relatively small $t_p$ we call influencers. From the simulation it is clear that not all actors with high link degree are influencers (see Fig.3). There are many actors with small degree in the group of the best 100 influencers.

Influencers in the case of 0-1-2 effect are not the same actors which have the smallest $t_p$ value without 0-1-2 effect (see Fig.4).

**DISSCUSION**

In according to 0-1-2 effect and as result of simulations data analysis it is clear that to be an influencer an actor not only has to have big number of followers, but these followers have to be linked between them. In social networks some "stars" have disconnected followers and as result they are not influencers.

In randomization process the Small World property is destroyed, but the stars continue to be stars as regarding to their degree. Simulation shows that influencers does not exist in this case (see Fig.2), this points that influencer can exist only in Small World networks because of conformity, which may be taking into account by 0-1-2 effect. An opinion has a limited time interval when it is interesting for crowd. Actually it means that the tipping point for network without Small World topology may be not reachable.

Known characteristics of an actor in a network can not indicate if he or she is a potential influencer. It's clear that an influencer must not have a low degree and must have a high clustering coefficient value. To become an influencer, a special position of an actor in the network is needed and this position is not a local property of the actor. Recently was proposed to use *k*-shell decomposition of the network for detection influencers (Kitsak at al. 2010). Further investigations will be concentrated on accurate definition of this position together with introducing of new topological metrics of a network.